\documentclass[11pt,a4paper]{article}
\usepackage{jheppub}
\usepackage{comment} 

\title{On the breaking of collinear factorization in QCD}
\author{Jeffrey R. Forshaw,}
\author{Michael H. Seymour,}
\author{Andrzej Si\'odmok}
\affiliation{Consortium for Fundamental Physics, School of Physics \& Astronomy, University of Manchester, Oxford Road, Manchester M13 9PL. U.K.}
\emailAdd{Jeffrey.Forshaw@manchester.ac.uk}
\emailAdd{Michael.Seymour@manchester.ac.uk}
\emailAdd{Andrzej.Siodmok@manchester.ac.uk}

\abstract{We investigate the breakdown of collinear factorization for
  non-inclusive observables in hadron-hadron collisions. For pure QCD
  processes, factorization is violated at the three-loop level and it
  has a structure identical to that encountered previously in the case
  of super-leading logarithms. In particular, it is driven by the
  non-commutation of Coulomb/Glauber gluon exchanges with other soft
  exchanges. Beyond QCD, factorization may be violated at the two-loop
  level provided that the hard subprocess contains matrix element
  contributions with phase differences between different colour
  topologies.}

\preprint{MAN/HEP/2012/05}
\usepackage[normalem]{ulem}

\begin{document}

\maketitle

\section{Introduction}

Factorization theorems are a cornerstone of QCD calculations. In
particular, it is the inclusive collinear factorization theorem that allows the
calculation of cross sections in hadron-hadron collisions as the
convolution of universal (i.e.\ process independent) parton distribution
functions and (perturbatively calculable) partonic cross sections.

The factorization theorem has been proven for
sufficiently inclusive observables in the final state of the scattering
of colourless hadrons \cite{Collins:1988ig}. However, it is often assumed that partonic scattering amplitudes also factorize, e.g. in the calculation of amplitudes to high orders and their resummation to all orders. Moreover, this assumed factorization forms the basis upon which parton shower Monte Carlo event generators are constructed. In neither of these cases is the factorization guaranteed. Indeed, in a recent paper, Catani,
de~Florian and Rodrigo  (CdFR) explicitly wrote down
amplitude and cross section contributions that  violate collinear
factorization for non-inclusive processes with two incoming coloured partons \cite{Catani:2011st}. 

In this paper we aim to shed light on the breakdown of collinear factorization. To that end, we re-consider the results presented in CdFR. Most of their
paper concerns the question of whether there are factorization-violating
terms in higher-order corrections to  QCD amplitudes. The answer, for initial-state (space-like, SL)
collinear branching in a process with two incoming coloured partons, is
definitely \emph{Yes}. CdFR derived an all-orders expression for the
singularities of the collinear splitting amplitude and gave its explicit
realization at one- and two-loop order. We discuss the structure of
these results and isolate the factorization violating terms in a way which makes it clear that their physical origin is the Coulomb interaction
between the two incoming partons long before the hard interaction or between two outgoing partons much later than the hard interaction.

CdFR then discuss whether these factorization violations at amplitude
level lead to factorization violation at cross section level. The answer
at one-loop order is definitely~\emph{No}. CdFR does however calculate a
contribution that violates cross section factorization at two-loop
order. We show below that this result is, in fact, only non-zero for
certain electroweak processes. For QCD scattering processes,
factorization violation effects cancel and
the answer at two-loop order remains \emph{No}. In the following, we argue
that this will not continue to be the case at still higher orders, and
\emph{we do expect factorization violation at three-loop order in the
  SL collinear limit in parton-parton cross sections.} We specify
the form we expect for these terms.

This paper is set out as follows. In the remainder of this introduction,
we briefly recall our work on the logarithmic
structure of high-order corrections to non-inclusive QCD observables such as
dijet production with a jet veto
(often referred to as `gaps between jets') \cite{Forshaw:2006fk,Forshaw:2008cq}. As we shall see, this is relevant
because the ``super-leading logarithms'' we identified there have the
identical structure and physical origin as the factorization violating
terms discussed here. Then, in section~\ref{general}, we recap some of the
formalism used by CdFR in deriving results in the SL collinear limit.
In section~\ref{one-loop} we review the results of CdFR for the one-loop
corrections to factorization of SL collinear emission at amplitude
level. We rewrite the final result in a form in which it is manifest
that the physical origin of the factorization violating term arises from an effective
Coulomb interaction between the two  incoming partons. We show that it does
not contribute to physical cross sections at this order. In
section~\ref{two-loop-amplitude} we consider the two-loop corrections at
amplitude level. We confirm the CdFR results and again rewrite them in a
form in which the connection with Coulomb gluons is manifest. In
section~\ref{two-loop-sigma} we show that the factorization violating
terms do not contribute to QCD cross sections at this order
either. We discuss the conditions under which they
could give a non-zero contribution and, within the Standard Model,
show that these are only satisfied by certain electroweak processes. Finally, in
section~\ref{conclusion} we
make some concluding remarks, including a prediction of the form of the term that will violate the cross section
factorization of collinear singularities at three-loop order
(for coloured incoming partons).

In
Refs.~\cite{Forshaw:2006fk,Forshaw:2008cq} we studied the
structure of the `gaps-between-jets' cross section in QCD. As is well known
(see for example the series of papers by Sterman and
collaborators \cite{Kidonakis:1998nf,Oderda:1998en,Berger:2001ns}),
``leading'' logarithmic terms
$\sim\alpha_s^n\log^n(Q/Q_0)$, where $Q$ is the hard scale and $Q_0$ the
veto scale, can be summed to all orders by a matrix-valued evolution in
colour space. The anomalous dimension matrix that drives this receives a
real (strictly, Hermitian) contribution from virtual gluons with
light-like momenta and in what follows we shall refer to these as `eikonal gluons'.
It also receives an imaginary (strictly, anti-Hermitian) contribution from space-like
virtual gluons that correspond to rescattering between the
partons long before or after the hard scattering process. We refer to these as `Coulomb gluons', although they are also  referred to as Glauber gluons in the literature.

In Ref.~\cite{Forshaw:2006fk}, we considered the first correction to
this picture. Specifically, we allowed for one additional gluon, either real or virtual, with momentum
corresponding to phase space regions lying outside the jet veto region\footnote{The virtual corrections in the anomalous dimension matrix are instead integrated over the region of phase space where real emissions are forbidden, i.e.\ inside the jet veto region.}.  Following the
%work of another of us, \cite{Dasgupta:2002bw}, we expected to find a
work of Dasgupta and Salam, \cite{Dasgupta:2002bw}, we expected to find a tower of additional leading logarithms, called non-global
logarithms. Physically these arise from the requirement that although
radiation outside the veto region is allowed, secondary radiation from it back
into the veto region is not. We did find these non-global logarithms. To
our surprise we also found that, even when the `out-of-gap' gluon is far
forward (i.e.\ collinear to an incoming parton), the subsequent colour evolution differs, depending
on whether the low-angle gluon is real or virtual. This triggers a breakdown of the `plus prescription' and to the conclusion that collinear logarithms can only be factorized into the incoming parton density functions at or below the scale $Q_0$. 
%Although this result was derived in
%the soft eikonal approximation, it depends only on the direction of the
%out-of-gap gluon and is equally valid for hard initial-state collinear
%radiation. 
In subsequent conference presentations \cite{conferences}, we
described this as a breakdown of QCD coherence and discussed further the fact
that it signals a breakdown of
collinear factorization for processes with incoming coloured partons. In
the gaps-between-jets calculation, this mis-cancellation means that the
integration over the momentum of the out-of-gap gluon is
double-logarithmic and leads to the appearance of super-leading
logarithms: one out-of-gap gluon contributes a tower
$\sim\alpha_s^n\log^{n+1}(Q/Q_0),\;n\ge4$, and we speculated that the
leading behaviour at high order is actually
$\sim\alpha_s^n\log^{2n-3}(Q/Q_0)$. This speculation was subsequently
confirmed, at $n=5$ with two out-of-gap gluons, in a fixed-order
calculation~\cite{Keates:2009dn}. Ref.~\cite{Forshaw:2008cq} re-performed
the calculation using the colour basis independent notation
\cite{Catani:1996vz} also used by CdFR and showed that it is a general
feature of QCD scattering amplitudes. These effects were also considered
for event shape variables in Ref.~\cite{Banfi:2010xy} where they were
called ``coherence violating logarithms'' since, depending on the class
of event shapes, they can be super-leading, leading or sub-leading. Importantly, these effects are generic to all non-inclusive final state observables (and they are certainly not unique to non-global observables).

As emphasized in Ref.~\cite{Forshaw:2008cq}, the origin of the mis-cancellation between the evolution of a system with
a real or virtual collinear gluon comes entirely from a mismatch between
the colour matrices of the Coulomb gluon contributions to the two
amplitudes. For inclusive observables, this would only lead to a phase (a purely
anti-Hermitian transformation in colour space) in the amplitude and
would not contribute to physical cross sections. However, the non-Abelian
nature of QCD means that the colour matrix of the Coulomb gluon
contribution does not commute with that of other soft (eikonal) gluon exchanges that generate
the Hermitian part of the evolution in the case of non-inclusive observables. This non-commutation can (and does)
lead to Coulomb-induced factorization violation in physical cross sections. This occurs first at fourth order relative to the
hard process, since we need one collinear emission, two Coulomb
exchanges (to give a real Hermitian contribution, $\sim(i\pi)^2)$ and one
real exchange (for them to not commute with).

In the following, we will show that the physical origin of the
non-factorizing terms identified by CdFR is identical to that of the
super-leading logarithms and that the same argument about their leading
effect on physical cross sections applies: they can contribute to fourth-order corrections to hard QCD processes, i.e.\ three-loop corrections to
the single-collinear splitting functions.

\section{Space-like collinear factorization}
\label{general}

At tree level, the factorization theorem for QCD amplitudes in the
multiparton collinear limit reads
\begin{equation}
\label{treefactorization}
  \,\Bigl|\, \mathcal{M}^{(0)} \Bigr\rangle
  \approx
  \boldsymbol{Sp}^{(0)}
  \,\Bigl|\, \overline{\mathcal{M}}^{(0)} \Bigr\rangle.
\end{equation}
$\mathcal{M}^{(0)}$ is the tree-level amplitude\footnote{Strictly
  speaking, we mean the lowest order non-zero amplitude, which for some
  processes could mean a loop amplitude, for example
  $gg\to\gamma g$.} for some process
involving $n$ coloured partons. We follow the convention of CdFR in
defining all momenta as outgoing, so incoming partons have $p_i^0<0$.
To be concrete, when discussing processes with two incoming partons (as
we are throughout this paper) we assume that partons 1 and $m\!+\!1$ are
incoming and all others are outgoing. $|\mathcal{M}^{(0)}\rangle$ is the
representation of $\mathcal{M}^{(0)}$ in colour space \cite{Catani:1996vz}. Equation~(\ref{treefactorization})
defines the factorization theorem of the amplitude $\mathcal{M}^{(0)}$
in the multicollinear limit in which $m$ of the partons become collinear
(in the SL case of interest, one of the $m$ is the incoming parton~1).
The accuracy is as specified in CdFR: in the multicollinear case
it gives the dominant contribution in which all scalar products among
the $m$ collinear partons are of the same order and all vanish together,
and does not include sub-dominant (but still singular) contributions. In
the particular case of the two-parton collinear limit there are no
sub-dominant corrections and this formula captures the entire singular
collinear behaviour. $\overline{\mathcal{M}}^{(0)}$ is the tree-level
amplitude for a process in which the $n\!-\!m$ non-collinear partons'
momenta and colours are the same as in $\mathcal{M}^{(0)}$, but the $m$
collinear partons are replaced by a single (on-shell) parton with momentum
\begin{equation}
  \widetilde{P}^\mu = p_{\mbox{\scriptsize jet}}^\mu -
  \frac{p_{\mbox{\scriptsize jet}}^2\,n^\mu}{2p_{\mbox{\scriptsize jet}}\!\cdot\!n}\,,
  \qquad p_{\mbox{\scriptsize jet}} = \sum_{i=1}^m p_i\,,
\end{equation}
where $n$ is a light-like ($n^2=0$) vector introduced to define the
collinear limit.
In the SL case of interest, $\widetilde{P}^0<0$.
Flavour conservation in the collinear limit defines the
flavour of parton $\widetilde{P}$ and colour conservation implies that
it has colour
\begin{equation}
  \label{colourPtilde}
  \boldsymbol{T}_{\widetilde{P}} = \sum_{i=1}^m \boldsymbol{T}_i =
  -\sum_{j=m+1}^n \boldsymbol{T}_j\,.
\end{equation}
Note that in general, colour conservation means that, when acting on
physical states, the colour operators obey
\begin{equation}
  \sum_{i=1}^n \boldsymbol{T}_i = 0\,,
\end{equation}
from which the second equality of Eq.~(\ref{colourPtilde}) follows.

$\boldsymbol{Sp}^{(0)}$ is an operator describing the collinear
splitting $\widetilde{P}(\boldsymbol{T}_{\widetilde{P}})\to
p_1(\boldsymbol{T}_1)+p_2(\boldsymbol{T}_2)+\ldots+p_m(\boldsymbol{T}_m)$. It can be represented by a
non-square matrix in colour space, since it acts on the space of $n\!-\!m\!+\!1$
partons to produce a state in the space of $n$ partons.
Equation~(\ref{treefactorization}) is defined as ``strict''
factorization, because $\boldsymbol{Sp}^{(0)}$ depends \emph{only} on the
colours and momenta of the $m$ collinear partons and
$\overline{\mathcal{M}}^{(0)}$ depends \emph{only} on the colours and
momenta of the non-collinear partons and $\widetilde{P}$.

CdFR show that the structure of Eq.~(\ref{treefactorization}) continues
to all orders:
\begin{equation}
\label{allorderfactorization}
  \,\Bigl|\, \mathcal{M} \Bigr\rangle
  \approx
  \boldsymbol{Sp}
  \,\Bigl|\, \overline{\mathcal{M}} \Bigr\rangle,
\end{equation}
where the kets $|\mathcal{M}\rangle$ and
$|\overline{\mathcal{M}}\rangle$ and the operator $\boldsymbol{Sp}$ have
perturbative (loop) expansions starting from 0~loops, i.e.
\begin{equation}
  \,\Bigl|\, \mathcal{M}^{(l)} \Bigr\rangle
  = \sum_{l'=0}^l
  \boldsymbol{Sp}^{(l')} \Bigl|\, \overline{\mathcal{M}}^{(l-l')}\Bigr\rangle~.
  \end{equation}  
Equation~(\ref{allorderfactorization}) is a
factorization, but it is described as a ``generalized'' factorization,
because although $\overline{\mathcal{M}}$ still depends only on the
colours and momenta of the non-collinear partons and $\widetilde{P}$,
$\boldsymbol{Sp}$ depends in general on both the collinear and non-collinear
partons.

In this paper we study the infrared poles (i.e.\ $\epsilon$ poles in
dimensional regularization) of the multi-collinear splitting matrix
$\boldsymbol{Sp}$ at various orders\footnote{Of course, taking $m=2$ we
  recover the two-parton collinear limit. As shown in CdFR, in this case
  it is possible to calculate the non-factorizing terms exactly in $d$
  dimensions but, while important for detailed calculations, we do not
  believe that this illustrates any additional physics, so we work in
  the multicollinear case throughout.}. In fact, the only ingredients
that we need to do this are Eq.~(\ref{allorderfactorization}) and the
results for the singularities of the on-shell amplitudes
$|\mathcal{M}^{(l)}\rangle$ and $|\overline{\mathcal{M}}^{(l)}\rangle$.
These have been known at 1~loop ($l=1$) for many years, and were first
written down in the colour basis independent notation in
Ref.~\cite{Catani:1996vz}. They were written down for the two-loop case in
Ref.~\cite{Catani:1998bh} and have recently been explored at higher loops
\cite{Dixon:2008gr,Becher:2009cu,Gardi:2009qi,Becher:2009qa,Dixon:2009ur}.
We define the singular parts of the $l$-loop amplitude as
\begin{equation}
  \,\Bigl|\, \mathcal{M}^{(l)} \Bigr\rangle
  = \sum_{l'=1}^l
  \boldsymbol{I}^{(l')}\,\Bigl|\, \mathcal{M}^{(l-l')} \Bigr\rangle
  +\,\Bigl|\, \mathcal{M}^{(l)\mbox{\scriptsize fin.}} \Bigr\rangle \,,
\end{equation}
with analogous definitions for operators $\overline{\boldsymbol{I}}^{(l')}$
for the reduced amplitude $|\overline{\mathcal{M}}^{(l)}\rangle$. We give
explicit expressions for these operators in the appropriate sections below.

We can use the factorization of amplitudes in the collinear
limit to derive the factorization of cross sections:
\begin{equation}
  \Bigl\langle \mathcal{M}
  \,\Bigl|\, \mathcal{M} \Bigr\rangle
  \approx
  \Bigl\langle \overline{\mathcal{M}}
  \,\Bigl|\, \mathbf{P}
  \,\Bigl|\, \overline{\mathcal{M}} \Bigr\rangle \,,
  \qquad
  \mathbf{P} = \boldsymbol{Sp}^\dagger \, \boldsymbol{Sp} \,.
\end{equation}
If $\boldsymbol{Sp}$ obeys strict factorization, then $\mathbf{P}$ does
too. On the contrary, it is possible that strict factorization is broken at
the amplitude level, but nevertheless the factorization-violating terms
do not contribute to physical cross sections. In fact this is true for
time-like (TL) collinear splitting, less trivially for SL collinear
splitting in processes with only one incoming coloured parton
(e.g.\ deep inelastic scattering) and, less trivially still, for
general processes at one-loop level and QCD processes at two-loop level,
as we will show.

We close this section by noting a property of QCD tree-level matrix
elements $|\overline{\mathcal{M}}^{(0)}\rangle$ (and also
$\boldsymbol{Sp}^{(0)}$) that we will make use of in the following. In
the colour basis independent notation, one works in an orthonormal basis
for the colour states of a given scattering process. It is natural (and
has been done in all calculations to date) to define these basis states
as combinations of generators, delta functions and structure constants
($if^{abc}$ and $d^{abc}$) with real coefficients. It was noticed in
Ref.~\cite{Seymour:2005ze} that, in these natural bases, all anomalous
dimension matrices calculated in the literature are (complex) symmetric
matrices. In Ref.~\cite{Seymour:2008xr} it was proved that this is
because the matrix representation of any colour operator
$\boldsymbol{T}_i\!\cdot\!\boldsymbol{T}_j$ (which is Hermitian in any
orthonormal basis) is real in these bases, and hence
symmetric. In the following we will exploit the existence of such a
basis, even if it is not necessary that it be specified explicitly.

The vector representation of the amplitude
$|\overline{\mathcal{M}}^{(0)}\rangle$ is given by its projections onto
the basis vectors of the space. An important point in the following is
the fact that tree-level QCD amplitudes do not introduce any additional
phases into this representation. That is, all the elements of the vector
representation of $|\overline{\mathcal{M}}^{(0)}\rangle$ have the same
phase and so the outer product $|\,\overline{\mathcal{M}}^{(0)}\rangle\,
\langle\overline{\mathcal{M}}^{(0)}|$  (which is called the ``hard matrix'' in the
language of Sterman et al.) can be represented as a matrix with real entries.
Since it is also Hermitian it must be symmetric.
The same argument applies to $\boldsymbol{Sp}^{(0)}$. It is a tree-level QCD operator whose matrix representation is a transformation between
two orthonormal spaces (of different dimensions). If both of those
spaces are represented by natural bases, the matrix representation of
$\boldsymbol{Sp}^{(0)}$ is real.

\section{Space-like collinear factorization at one loop}
\label{one-loop}

At one loop \cite{Catani:1996vz}, the operator $\boldsymbol{I}$
can be written\footnote{To avoid a cluttered notation, we do not explicitly indicate
  that $\boldsymbol{I}$, and all the related operators we consider, are
  renormalized and only have poles of IR origin. Thus we drop the $R$
  superscript used by CdFR, as well as the argument of $\alpha_S$, which
  is the $\overline{\mbox{MS}}$ coupling evaluated at $\mu^2$. Finally,
  we also drop the argument of $\boldsymbol{I}(\epsilon)$, except where
  it is not $\epsilon$, when we explicitly write it.}
\begin{equation}
  \boldsymbol{I}^{(1)}=\frac{\alpha_s}{2\pi}\,\frac12\left\{
  -\sum_{i=1}^n\left(\frac1{\epsilon^2}C_i
  +\frac1\epsilon\gamma_i\right)
  -\frac1\epsilon\sum_{i,j=1\atop i\not=j}^n
  \boldsymbol{T}_i\cdot\boldsymbol{T}_j\ln\left(\frac{-s_{ij}-i0}{\mu^2}\right)
  \right\},
\end{equation}
with an exactly analogous expression for $\overline{\boldsymbol{I}}^{(1)}$.
Examining the analytic structure of the logarithm, we see that when $i$
and $j$ are both incoming partons, or both outgoing partons, it
contributes an imaginary part. Physically, this imaginary part
corresponds to an absorptive
contribution coming from on-shell parton-parton scattering long before,
or long after, the hard scattering process. One can rewrite this
equation to make the imaginary part manifest, for example as
\begin{equation}
  \label{I1rewritten}
  \boldsymbol{I}^{(1)}=\frac{\alpha_s}{2\pi}\,\frac12\left\{
  -\sum_{i=1}^n\left(\frac1{\epsilon^2}C_i
  +\frac1\epsilon\gamma_i
  +\frac{i\pi}\epsilon C_i\right)
  -\frac1\epsilon\sum_{i,j=1\atop i\not=j}^n
  \boldsymbol{T}_i\cdot\boldsymbol{T}_j\ln\left(\frac{|s_{ij}|}{\mu^2}\right)
  + \frac{2i\pi}{\epsilon} \boldsymbol{T}_s^2
  \right\},
\end{equation}
where
\begin{equation}
  \boldsymbol{T}_s = \boldsymbol{T}_1+\boldsymbol{T}_{m+1}
  =-\!\!\sum_{i\not=1,m+1}^n\boldsymbol{T}_i
\end{equation}
is the total colour in the $s$~channel. In this form, one
can see that the contribution comes symmetrically from initial-state and
final-state interactions, which, since colour is conserved, are
equal. It is more useful for practical calculation to write this
entirely in terms of the initial-state contribution. Moreover, purely
Abelian imaginary parts $\sim i\pi C_i$ cannot contribute to cross
sections. Therefore, in the following, we shall write the physically
important imaginary part as
$\frac{4i\pi}{\epsilon} \boldsymbol{T}_1 \cdot \boldsymbol{T}_{m+1}$.

Writing the operator as in Eq.~(\ref{I1rewritten}), we are separating
different regions of the integration over gluon loop momenta:
``eikonal'' gluons (i.e.~with on-shell momenta) give rise to the
logarithm, whilst ``Coulomb'' gluons (i.e.~with space-like momenta)
correspond to on-shell external parton-parton scattering and give rise
to the $i \pi$ term.
It will turn out that the factorization breaking contributions can be understood entirely in terms of the non-trivial colour operators associated with Coulomb and eikonal gluon exchanges.

The singularities of the one-loop splitting matrix, defined by 
\begin{equation}
  \boldsymbol{Sp}^{(1)}
  =   \boldsymbol{I}_C^{(1)} \, \boldsymbol{Sp}^{(0)}
  + \boldsymbol{Sp}^{(1)\mbox{\scriptsize fin.}}~,
\end{equation}
can now be extracted using
\begin{equation}
\boldsymbol{I}_C^{(1)}\boldsymbol{Sp}^{(0)} = \boldsymbol{I}^{(1)}\boldsymbol{Sp}^{(0)} - \boldsymbol{Sp}^{(0)}\overline{\boldsymbol{I}}^{(1)}~,
\end{equation}
which we write as\footnote{See the discussion above Eq.~(5.49) in CdFR.
}
\begin{equation}
  \boldsymbol{I}_C^{(1)} = \boldsymbol{I}^{(1)} - \overline{\boldsymbol{I}}^{(1)}.
\end{equation}
After some colour algebra, including the colour conservation condition
in Eq.~(\ref{colourPtilde}), we obtain
\begin{eqnarray}
  \label{Ic1}
\!\!
  \boldsymbol{I}_C^{(1)}&=&\frac{\alpha_s}{2\pi}\,\frac12
  \left\{
  \left(\frac1{\epsilon^2}C_{\widetilde{P}}+\frac1\epsilon\gamma_{\widetilde{P}}\right)
  -\sum_{i=1\atop{}}^m
  \left(\frac1{\epsilon^2}C_i+\frac1\epsilon\gamma_i
  -\frac2\epsilon C_i\ln|z_i|\right)
  -\frac{i\pi}{\epsilon}\left(C_{\widetilde{P}}-C_1+\sum_{i=2}^mC_i\right)
\!\!\!
  \right.\nonumber\\&&\left.\hspace*{10em}
  {}- \frac1\epsilon \sum_{i,\ell=1\atop i\not=\ell}^m
  \boldsymbol{T}_i\cdot\boldsymbol{T}_\ell\ln\frac{|s_{i\ell}|}{|z_i|\,|z_\ell|\,\mu^2}
    \right\}+\widetilde{\mathbf{\Delta}}_C^{(1)},
\end{eqnarray}
with
\begin{equation}
  \label{deltaC1}
  \widetilde{\mathbf{\Delta}}_C^{(1)} =
  \frac{\alpha_s}{2\pi}\,\left\{2\times\frac{i\pi}{\epsilon}
  \boldsymbol{T}_{m\!+\!1}\cdot\left(\boldsymbol{T}_1-\boldsymbol{T}_{\widetilde{P}}\right)
  \right\}.
\end{equation}
All notation not explicitly defined here is taken over directly from
CdFR, but in brief: $C_i$ is the Casimir of particle type $i$,
$\gamma_i$ is a flavour dependent real constant and
$z_i=p_i\!\cdot\!n/\widetilde{P}\!\cdot\!n$ is positive for the incoming
collinear parton ($i=1$) and negative for the outgoing collinear partons
($i=2 \ldots m$). 

Note that Eq.~(\ref{Ic1}) is exactly equal to CdFR Eq.~(5.31), but it is
rewritten in such a way that our $\widetilde{\mathbf{\Delta}}_C^{(1)}$
is different to their $\mathbf{\Delta}_{mC}^{(1)}$ in Eq.~(5.32). The
terms explicitly written in Eq.~(\ref{Ic1}) obey strict factorization:
they depend only on the kinematics and colours of the collinear
partons. On the other hand, $\widetilde{\mathbf{\Delta}}_C^{(1)}$
 violates strict factorization since it depends on the colour of a
non-collinear parton, i.e.~the other incoming parton, $m\!+\!1$. Written in
this form, it is very clear that the violation of strict factorization
observed by CdFR is directly related to a mismatch between the colour
structures of the Coulomb gluon contributions in the full
($\boldsymbol{T}_{m\!+\!1}\cdot\boldsymbol{T}_1$) and factorized
($\boldsymbol{T}_{m\!+\!1}\cdot\boldsymbol{T}_{\widetilde{P}}$) matrix elements. All
other terms obey strict factorization. This mismatch was illustrated diagrammatically in Figure 1 of Ref.~\cite{Forshaw:2008cq} and in the associated discussion.

Turning to the factorization theorem for cross sections, we have
\begin{eqnarray}
  \mathbf{P}^{(1)} &=& \boldsymbol{Sp}^{(0)\dagger} \, \boldsymbol{Sp}^{(1)} +
  \boldsymbol{Sp}^{(1)\dagger} \, \boldsymbol{Sp}^{(0)}
  =
  \boldsymbol{Sp}^{(0)\dagger}\left(\boldsymbol{I}_C^{(1)}+\boldsymbol{I}_C^{(1)\dagger}
  \right)\boldsymbol{Sp}^{(0)}
  +\mathbf{P}^{(1)\mbox{\scriptsize fin.}}
\\&\equiv&   \boldsymbol{Sp}^{(0)\dagger}\boldsymbol{I}_P^{(1)}\boldsymbol{Sp}^{(0)}
  +\mathbf{P}^{(1)\mbox{\scriptsize fin.}},
\end{eqnarray}
\begin{equation}
\hspace*{-1em}
  \boldsymbol{I}_P^{(1)}=\frac{\alpha_s}{2\pi}
  \left\{
  \left(\frac1{\epsilon^2}C_{\widetilde{P}}+\frac1\epsilon\gamma_{\widetilde{P}}\right)
  -\sum_{i=1\atop{}}^m
  \left(\frac1{\epsilon^2}C_i+\frac1\epsilon\gamma_i
  -\frac2\epsilon C_i\ln|z_i|\right)
  - \frac1\epsilon \sum_{i,\ell=1\atop i\not=\ell}^m
  \boldsymbol{T}_i\cdot\boldsymbol{T}_\ell\ln\frac{|s_{i\ell}|}{|z_i|\,|z_\ell|\,\mu^2}
    \right\}.
\hspace*{-1em}
\end{equation}
We see that the divergent terms do now have a factorized structure. This is because
$\widetilde{\mathbf{\Delta}}_C^{(1)}$, the non-factorizing term in
$\boldsymbol{I}_C^{(1)}$, is anti-Hermitian and it has cancelled in the
combination $\boldsymbol{I}_C^{(1)}+\boldsymbol{I}_C^{(1)\dagger}$. The only
possible source of factorization violation is in the finite term,
\begin{equation}
  \mathbf{P}^{(1)\mbox{\scriptsize fin.}} = \boldsymbol{Sp}^{(0)\dagger} \,
  \boldsymbol{Sp}^{(1)\mbox{\scriptsize fin.}}
  +\mbox{h.c}\,,
\end{equation}
since in the multi-parton collinear limit it is not (yet) proven that
$\boldsymbol{Sp}^{(1)\mbox{\scriptsize fin.}}$ contains only (finite) factorizing terms. In the two-parton limit on the other hand,
$\boldsymbol{Sp}^{(1)\mbox{\scriptsize fin.}}$ has the same structure as
$\boldsymbol{Sp}^{(0)}$ and hence factorizes.

In summary, in this section, we have presented results that completely agree
with those of CdFR, but we believe that our writing of the
non-factorizing contribution to $\boldsymbol{I}_C^{(1)}$,
Eq.~(\ref{deltaC1}), displays its physical origin more clearly.

\section{Space-like collinear factorization at two loops: amplitude level}
\label{two-loop-amplitude}

At two loops \cite{Catani:1998bh}, the operator $\boldsymbol{I}$ is
\begin{equation}
  \boldsymbol{I}^{(2)}=-\frac12\left[\boldsymbol{I}^{(1)}\right]^2
  +\frac{\alpha_s}{2\pi}\left\{
  +\frac1\epsilon\,b_0\left[\boldsymbol{I}^{(1)}(2\epsilon)-\boldsymbol{I}^{(1)}
    \right]+K\,\boldsymbol{I}^{(1)}(2\epsilon)\right\}
  +\left(\frac{\alpha_s}{2\pi}\right)^2\frac1\epsilon
  \sum_{i=1}^nH_i^{(2)}\,,
\end{equation}
with an exactly analogous expression for
$\overline{\boldsymbol{I}}^{(2)}$. $K$ and $H_i^{(2)}$ are more (real)
constants that may be found in CdFR. Recall that we suppress the
$\epsilon$ argument of $\boldsymbol{I}^{(l)}$,
except where its argument is not $\epsilon$. 

The singularities of the two-loop splitting operator can then be extracted
as
\begin{equation}
  \boldsymbol{Sp}^{(2)}=
  \boldsymbol{I}_C^{(2)}\boldsymbol{Sp}^{(0)}
  +\boldsymbol{I}_C^{(1)}\boldsymbol{Sp}^{(1)}
  +\overline{\boldsymbol{Sp}}^{(2)\mbox{\scriptsize div.}}
  +\mbox{finite terms},
\end{equation}
where
\begin{equation}
  \overline{\boldsymbol{Sp}}^{(2)\mbox{\scriptsize div.}}=
  \left[\overline{\boldsymbol{I}}^{(1)},\boldsymbol{Sp}^{(1)\mbox{\scriptsize fin.}}\right],
\end{equation}
which vanishes if there are no factorization violating terms in
$\boldsymbol{Sp}^{(1)\mbox{\scriptsize fin.}}$.

We can extract $\boldsymbol{I}_C^{(2)}$ as
\begin{equation}
  \boldsymbol{I}_C^{(2)}=
  \boldsymbol{I}^{(2)}
  -\overline{\boldsymbol{I}}^{(2)}
  +\overline{\boldsymbol{I}}^{(1)}\left(\boldsymbol{I}^{(1)}
  -\overline{\boldsymbol{I}}^{(1)}\right),
\end{equation}
i.e.
\begin{eqnarray}
\!
  \boldsymbol{I}_C^{(2)} &=&
  -\frac12\left[\boldsymbol{I}_C^{(1)}\right]^2
  +\frac{\alpha_s}{2\pi}\left\{+\frac1\epsilon\,b_0\left[
  \boldsymbol{I}_C^{(1)}(2\epsilon)-\boldsymbol{I}_C^{(1)}\right]
  +K\,\boldsymbol{I}_C^{(1)}(2\epsilon)
  +\frac{\alpha_s}{2\pi}\,\frac1\epsilon
  \left(\sum_{i\in C}H_i^{(2)}-H_{\widetilde{P}}^{(2)}\right)
  \right\}
\!
\nonumber\\&&
   +\frac12\left[\overline{\boldsymbol{I}}^{(1)},\boldsymbol{I}_C^{(1)}\right].
\end{eqnarray}
Note that all of the parts that are non-trivial colour operators are
determined by one-loop operators. Thus, at two-loop order, the
violations of factorization can be understood in terms of one-loop
violations. To this end, it is useful to write
\begin{equation}
  \boldsymbol{I}_C^{(1)}=\boldsymbol{I}_C^{(1)\mbox{\scriptsize fact.}}+
  \widetilde{\mathbf{\Delta}}_C^{(1)}.
\end{equation}
$\boldsymbol{I}_C^{(1)\mbox{\scriptsize fact.}}$ depends only on the
collinear partons' colours and momenta and has both Hermitian and
anti-Hermitian parts, while $\widetilde{\mathbf{\Delta}}_C^{(1)}$
depends on the colours of collinear and non-collinear partons and is
purely anti-Hermitian. We then have
\begin{eqnarray}
  \boldsymbol{I}_C^{(2)\mbox{\scriptsize fact.}} &=&
  -\frac12\left[\boldsymbol{I}_C^{(1)\mbox{\scriptsize fact.}}\right]^2
  +\frac{\alpha_s}{2\pi}\left\{+\frac1\epsilon\,b_0\left[
  \boldsymbol{I}_C^{(1)\mbox{\scriptsize fact.}}(2\epsilon)-\boldsymbol{I}_C^{(1)\mbox{\scriptsize fact.}}\right]
  +K\,\boldsymbol{I}_C^{(1)\mbox{\scriptsize fact.}}(2\epsilon)
\phantom{\sum_{i\in C}}
\right.\nonumber\\&&\left.
  +\frac{\alpha_s}{2\pi}\,\frac1\epsilon
  \left(\sum_{i\in C}H_i^{(2)}-H_{\widetilde{P}}^{(2)}\right)
  \right\},
\end{eqnarray}
and a factorization violating contribution to
$\boldsymbol{I}_C^{(2)}$ of
\begin{eqnarray}
  \widetilde{\mathbf{\Delta}}_C^{(2)} &\equiv&
  \boldsymbol{I}_C^{(2)}
  -\boldsymbol{I}_C^{(2)\mbox{\scriptsize fact.}}
\\&=&
  \label{deltaC2}
  -\frac12\left\{\boldsymbol{I}_C^{(1)\mbox{\scriptsize fact.}}\widetilde{\mathbf{\Delta}}_C^{(1)}
  +\widetilde{\mathbf{\Delta}}_C^{(1)}\boldsymbol{I}_C^{(1)\mbox{\scriptsize fact.}}
  +\left(\widetilde{\mathbf{\Delta}}_C^{(1)}\right)^2\right\}
  +\frac12\left[\overline{\boldsymbol{I}}^{(1)},\widetilde{\mathbf{\Delta}}_C^{(1)}\right]
\nonumber\\&&
  +\frac{\alpha_s}{2\pi}\left\{+\frac1\epsilon\,b_0\left[
  \widetilde{\mathbf{\Delta}}_C^{(1)}(2\epsilon)-\widetilde{\mathbf{\Delta}}_C^{(1)}\right]
  +K\,\widetilde{\mathbf{\Delta}}_C^{(1)}(2\epsilon)
  \right\}
.
\end{eqnarray}
Note that, unlike $\widetilde{\mathbf{\Delta}}_C^{(1)}$, which is purely
anti-Hermitian, $\widetilde{\mathbf{\Delta}}_C^{(2)}$ contains both
Hermitian and anti-Hermitian parts.

In summary, in this section, we have shown results that completely agree
with those of CdFR, but again they are slightly rewritten to isolate
more clearly the physical origin of the non-factorizing contributions to
$\boldsymbol{I}_C^{(2)}$: they arise entirely from non-factorizing
contributions that are already present in the one-loop results.

\section{Space-like collinear factorization at two loops: cross section level}
\label{two-loop-sigma}

In order to compute cross sections, we need
\begin{eqnarray}
  \mathbf{P}^{(2)} &=& \boldsymbol{Sp}^{(0)\dagger} \, \boldsymbol{Sp}^{(2)} +
  \boldsymbol{Sp}^{(1)\dagger} \, \boldsymbol{Sp}^{(1)} +
  \boldsymbol{Sp}^{(2)\dagger} \, \boldsymbol{Sp}^{(0)}
\\&\equiv&
  \mathbf{P}^{(2)}_{\mbox{\scriptsize f.}}+\mathbf{P}^{(2)}_{\mbox{\scriptsize n.f.}}
  +\mbox{finite terms},
\end{eqnarray}
with a factorizing contribution of
\begin{equation}
  \mathbf{P}^{(2)}_{\mbox{\scriptsize f.}}=
  \boldsymbol{Sp}^{(0)\dagger}\left\{
  \boldsymbol{I}_C^{(2)\mbox{\scriptsize fact.}}
  +\left[\boldsymbol{I}_C^{(1)\mbox{\scriptsize fact.}}\right]^2
  +\boldsymbol{I}_C^{(2)\mbox{\scriptsize fact.}\dagger}
  +\left[\boldsymbol{I}_C^{(1)\mbox{\scriptsize fact.}\dagger}\right]^2
  +\boldsymbol{I}_C^{(1)\mbox{\scriptsize fact.}\dagger}\boldsymbol{I}_C^{(1)\mbox{\scriptsize fact.}}\right\}
  \boldsymbol{Sp}^{(0)}.
\end{equation}
Any factorization violation at two-loop order arises from the remainder:
\begin{eqnarray}
  \label{P2nf}
  \mathbf{P}^{(2)}_{\mbox{\scriptsize n.f.}}&=&
  \boldsymbol{Sp}^{(0)\dagger}\left\{\widetilde{\mathbf{\Delta}}_C^{(2)} +
  \widetilde{\mathbf{\Delta}}_C^{(2)\dagger} +
  \left(\widetilde{\mathbf{\Delta}}_C^{(1)}\right)^2
  - \widetilde{\mathbf{\Delta}}_C^{(1)}\boldsymbol{I}_C^{(1)\mbox{\scriptsize fact.}\dagger} + 
 \boldsymbol{I}_C^{(1)\mbox{\scriptsize fact.}}\widetilde{\mathbf{\Delta}}_C^{(1)}
 \right\}\boldsymbol{Sp}^{(0)}
\nonumber\\&&
  +\boldsymbol{Sp}^{(0)\dagger}\left\{\boldsymbol{I}_P^{(1)}\boldsymbol{Sp}^{(1)\mbox{\scriptsize fin.}}
  +\overline{\boldsymbol{Sp}}^{(2)\mbox{\scriptsize div.}}\right\}
  +\mbox{h.c.}
  +\mbox{finite terms}.
\end{eqnarray}
The first line contains all of the IR divergences associated with
divergent operators acting on the tree-level splitting operator, while
the second line contains possibly non-factorizing divergences due to
non-factorizing finite parts of the one-loop splitting operator (recall
that $\overline{\boldsymbol{Sp}}^{(2)\mbox{\scriptsize div.}}$ depends on
$\boldsymbol{Sp}^{(1)\mbox{\scriptsize fin.}}$).

Inserting the expression in Eq.~(\ref{deltaC2}) into the first line of
Eq.~(\ref{P2nf}), we obtain
\begin{equation}
  \label{zero}
  \mathbf{P}^{(2)}_{\mbox{\scriptsize n.f.}}=
  \frac12\boldsymbol{Sp}^{(0)\dagger}\left[\left(\overline{\boldsymbol{I}}^{(1)} +
    \overline{\boldsymbol{I}}^{(1)\dagger} +
    \boldsymbol{I}_C^{(1)\mbox{\scriptsize fact.}} +
    \boldsymbol{I}_C^{(1)\mbox{\scriptsize fact.}\dagger}
    \right),
    \widetilde{\mathbf{\Delta}}_C^{(1)}
    \right]\boldsymbol{Sp}^{(0)}
 +\mbox{second line}.
\end{equation}
This important expression has a very simple physical interpretation. It expresses the fact that Coulomb exchange (which is the physics of $\widetilde{\mathbf{\Delta}}_C^{(1)}$) does not commute with eikonal exchanges between any two partons in the matrix element (which is the only relevant physics of $\overline{\boldsymbol{I}}^{(1)}$ and $\boldsymbol{I}_C^{(1)}$ since the Coulomb exchanges in these operators cancel after adding the Hermitian conjugate\footnote{For that reason the superscript `fact.' is actually redundant in Eq.~(\ref{zero}).}).

We can relate this commutator to that which is explicit in the
derivation of super-leading logarithms in Ref.~\cite{Forshaw:2008cq} by considering the limit in which the partons of the reduced matrix element divide into two clusters that are separated by a large interval in rapidity,~$Y$. In this case, the only large $s_{ij}$ are associated with the case where partons $i$ and $j$ are on opposite sides of the interval. Taking the leading~$Y$ behaviour is therefore equivalent to neglecting $\boldsymbol{I}_C^{(1)}$ and simplifying
\begin{equation} 
\overline{\boldsymbol{I}}^{(1)} + \overline{\boldsymbol{I}}^{(1)\dagger} \approx \frac{\alpha_s}{2 \pi}\,\frac{2}{\epsilon}\, Y \boldsymbol{T}_t^2~,
\end{equation}
where $\boldsymbol{T}_t$ is the colour exchanged across the rapidity interval.
\pagebreak[3]
Equation (\ref{zero}) then simplifies to
\begin{equation}
  \label{important}
  \mathbf{P}^{(2)}_{\mbox{\scriptsize n.f.}}\propto
  i\pi Y\boldsymbol{Sp}^{(0)\dagger}\left[\boldsymbol{T}_t^2,
  \boldsymbol{T}_{m\!+\!1}\cdot (\boldsymbol{T}_1-\boldsymbol{T}_{\widetilde{P}})
    \right]\boldsymbol{Sp}^{(0)}.
\end{equation}
This equation embodies exactly the same physical result as Eq.~(4.13) of
Ref.~\cite{Forshaw:2008cq}: the violation of factorization is driven by
the non-vanishing commutator of the eikonal gluon contribution with the
mismatch between the Coulomb gluon contributions of the full and
factorized matrix elements. To make this more precise, we should
explain the connection between the notation used here and that used in
Ref.~\cite{Forshaw:2008cq}. In that paper, the operator $\mathbf{t}_1^a$
is the colour part of $\boldsymbol{Sp}^{(0)}$,
$\mathbf{T}_1\cdot\mathbf{T}_2$ is $\boldsymbol{T}_{m\!+\!1}\cdot
\boldsymbol{T}_1$ and $\mathbf{t}_1\cdot\mathbf{t}_2$ is
$\boldsymbol{T}_{m\!+\!1}\cdot \boldsymbol{T}_{\widetilde{P}}$. Finally,
since $\boldsymbol{T}_{m\!+\!1}\cdot \boldsymbol{T}_{\widetilde{P}}$ and
$\boldsymbol{T}_t^2$ commute with $\boldsymbol{Sp}^{(0)}$ (in the sense
defined more precisely in Eq.~(4.12) of Ref.~\cite{Forshaw:2008cq}),
Eq.~(\ref{important}) can be written identically to Eq.~(4.13) of
Ref.~\cite{Forshaw:2008cq}.

In general, the commutator in Eq.~(\ref{zero}) is non-zero. However, we
can exploit its properties as a matrix in colour space to show that its
expectation value on a QCD tree amplitude
$\boldsymbol{Sp}^{(0)}|\overline{\mathcal{M}}^{(0)}\rangle$ is zero. As
discussed at the end of section~\ref{general}, there exists a basis in which the
matrix representation of any colour operator
$\boldsymbol{T}_i\!\cdot\!\boldsymbol{T}_j$ is real. Therefore, in such a
basis, the commutator in Eq.~(\ref{zero}) is purely imaginary. Since it
is also Hermitian, it must be antisymmetric. As we have  noted,
in such a basis QCD tree amplitudes are real and hence
$|\mathcal{M}^{(0)}\rangle\,\langle\mathcal{M}^{(0)}|$ is symmetric. The
final step of the argument is to note that
$\langle\mathcal{M}^{(0)}|A|\mathcal{M}^{(0)}\rangle$ can be written as
$\mathrm{Tr}\Bigl[\bigl(|\mathcal{M}^{(0)}\rangle\,\langle\mathcal{M}^{(0)}|\bigr)A\Bigr]$
and 
\begin{equation}
  \mathrm{Tr}\Bigl[SA\Bigr]=0
\end{equation}
for any symmetric, $S$, and antisymmetric, $A$, matrices. Since there
exists a basis in which the result is zero it must be zero in any basis.

Therefore, although the commutator in Eq.~(\ref{zero}) is non-zero, its
expectation value on a QCD tree amplitude is zero and
\begin{eqnarray}
  \mathbf{P}^{(2)}_{\mbox{\scriptsize n.f.}}&=&
  \mbox{second line}
  \qquad\mbox{(pure QCD processes)}.
\end{eqnarray}
We recall that the terms in the ``second line'' of Eq.~(\ref{P2nf})
arise from divergent one-loop corrections to finite one-loop corrections
to the splitting operator, which are not yet proven to factorize. We
have shown that, apart from this possible source, \emph{factorization
  violating terms in the one- and two-loop splitting amplitudes do
  \emph{not} contribute to factorization violation of QCD cross sections
  at the one- or two-loop level}.

CdFR considered the most
general case, in which it is true that the expectation value of the
non-factorizing term is non-zero, and did not consider the specific case
of QCD tree amplitudes. As mentioned in Section~\ref{general}, it is the
fact that the Feynman rules of QCD do not introduce any phases, so all
of the elements of the colour vector
$\boldsymbol{Sp}^{(0)}|\overline{\mathcal{M}}^{(0)}\rangle$ have the
same phase, that leads to this result. In order to have a non-zero
result, one must have a hard process with more than one colour flow,
with non-trivial phase differences between them. Within the Standard
Model, the only $2\to2$ processes that satisfy these conditions are
$q\bar{q}'\to q\bar{q}'$ and $q\bar{q}\to q\bar{q}$ in which an
$s$-channel colour singlet propagator ($W$ and $Z$ respectively)
introduces a non-trivial phase, which can then
interfere with $t$-channel octet (gluon) or singlet ($\gamma$ or $Z$)
exchange.

\section{Conclusion and discussion}
\label{conclusion}

We
have explored the results presented in CdFR and rewritten them in a way
that makes clear that the origin of the breakdown in collinear factorization  is a mismatch in the non-Abelian colour
matrix for Coulomb gluon exchange between the two incoming partons in
the full matrix element and the sub-process matrix element it is
factorized into, at one loop. At two-loop level, no fundamentally new
non-factorizing effects enter, and the two-loop violation of strict
factorization at the amplitude level is entirely determined by the
one-loop one. This mechanism is identical to the one studied in
Refs.~\cite{Forshaw:2006fk,Forshaw:2008cq}.

At one-loop level the violation of factorization in the amplitude is
anti-Hermitian and cannot contribute to physical cross sections. At
two-loop level the sum of all contributions corresponding to two Coulomb
gluon exchanges (in the amplitude, the conjugate amplitude and one in
each) similarly cancels. However, a non-zero commutator between one
Coulomb gluon exchange and one eikonal exchange gives a
Hermitian term that may  contribute to physical cross
sections. We find that this contribution does vanish for pure QCD processes. Again, this mirrors the conclusions of
Refs.~\cite{Forshaw:2006fk,Forshaw:2008cq}. Ref.~\cite{Forshaw:2008cq}
in particular has a discussion of the cancellation of these
terms in physical cross sections.

In Refs.~\cite{Forshaw:2006fk,Forshaw:2008cq}, we concluded that the
non-cancellation of Coulomb gluon effects does have a physical effect on
non-inclusive QCD observables at the three-loop level. We showed that this comes about because
only by exchanging two Coulomb gluons can we obtain a non-zero effect on
physical cross sections and only via a non-zero commutator between these
gluons and a further eikonal gluon can we avoid their complete
cancellation as a pure phase effect. This then induced a
mis-cancellation between the real and virtual initial-state collinear
emissions (i.e.\ the breakdown of the `plus-prescription'), which resulted in a ``super-leading'' logarithm.

The all-order development of CdFR can be
used at three loops to produce the results for the operator
$\boldsymbol{I}_C^{(3)}$, its non-factorizing part
$\mathbf{\Delta}_C^{(3)}$, and its effect on physical cross
sections~$\mathbf{P}^{(3)}_{\mbox{\scriptsize n.f.}}$. Exactly the
mechanism just
described does indeed occur: the commutator between two non-factorizing
one-loop Coulomb contributions and one factorizing eikonal contribution
produces a non-factorizing term in the cross section that does not
cancel. In order to display the results, we define one additional piece
of notation: the Coulomb gluon parts of $\boldsymbol{I}^{(1)}$ and
$\overline{\boldsymbol{I}}^{(1)}$ are defined as
$\widetilde{\mathbf{\Delta}}_1^{(1)}$ and
$\widetilde{\mathbf{\Delta}}_{\widetilde{P}}^{(1)}$ respectively (and
therefore
$\widetilde{\mathbf{\Delta}}_C^{(1)}=\widetilde{\mathbf{\Delta}}_1^{(1)}-\widetilde{\mathbf{\Delta}}_{\widetilde{P}}^{(1)}$). We
obtain terms that are in one-to-one
correspondence with those obtained for the first non-zero super-leading
logarithm in Ref.~\cite{Forshaw:2008cq}. In the 2-parton collinear
limit, in which the eikonal part of $\boldsymbol{I}_C^{(1)}$ is a pure
number, we obtain
\begin{eqnarray}
  \label{3loopguess}
  \mathbf{P}^{(3)}_{\mbox{\scriptsize n.f.}}&\sim&\phantom{+}
  \frac16
  \boldsymbol{Sp}^{(0)\dagger}\left(
  \left[\widetilde{\mathbf{\Delta}}_1^{(1)}
  ,\left[\widetilde{\mathbf{\Delta}}_1^{(1)},\overline{\boldsymbol{I}}^{(1)}+\overline{\boldsymbol{I}}^{(1)\dagger}\right]
  \right]
  -
  \left[\widetilde{\mathbf{\Delta}}_{\widetilde{P}}^{(1)}
  ,\left[\widetilde{\mathbf{\Delta}}_{\widetilde{P}}^{(1)},\overline{\boldsymbol{I}}^{(1)}+\overline{\boldsymbol{I}}^{(1)\dagger}\right]
  \right]
\right)
  \boldsymbol{Sp}^{(0)}
\nonumber\\&&
  +\frac12
  \boldsymbol{Sp}^{(0)\dagger}\left(
  \left[\widetilde{\mathbf{\Delta}}_{\widetilde{P}}^{(1)}
  ,\left[\widetilde{\mathbf{\Delta}}_{\widetilde{P}}^{(1)},\overline{\boldsymbol{I}}^{(1)}+\overline{\boldsymbol{I}}^{(1)\dagger}\right]
  \right]
  -
  \left[\widetilde{\mathbf{\Delta}}_{\widetilde{P}}^{(1)}
  ,\left[\widetilde{\mathbf{\Delta}}_1^{(1)},\overline{\boldsymbol{I}}^{(1)}+\overline{\boldsymbol{I}}^{(1)\dagger}\right]
  \right]
\right)
  \boldsymbol{Sp}^{(0)}.
\phantom{(9.9)}
\end{eqnarray}
The structure of the first line is identical to the contribution to
super-leading logarithms from configurations in which the out-of-gap
gluon has the highest $k_\perp$ (see Eq.~(4.18) of
Ref.~\cite{Forshaw:2008cq}) and is driven by the mismatch between the
double-commutator of one eikonal exchange and two Coulomb exchanges in
the full and factorized matrix element. The structure and relative
normalization of the second line
are identical to the contribution to super-leading logarithms from
configurations in which the out-of-gap gluon has the second-highest
$k_\perp$ (see Eq.~(4.20)\footnote{This is Eq.~(4.19) in the
  \texttt{arxiv} version of the paper.} of Ref.~\cite{Forshaw:2008cq})
and is driven by the mismatch between the commutator of one eikonal
exchange and one Coulomb exchange in the full and factorized matrix
element, with a second Coulomb exchange in the factorized matrix element
giving an overall real Hermitian contribution. To obtain the result in
the multi-parton collinear limit, one simply replaces
$\overline{\boldsymbol{I}}^{(1)}$ by
$\overline{\boldsymbol{I}}^{(1)}+\boldsymbol{I}_C^{(1)}$.

We also note that the non-vanishing double-commutators in
Eq.~(\ref{3loopguess}) are identical to those presented in
Ref.~\cite{DelDuca:2011ae} (Eq.~(4.10)), where they triggered the
failure of gluon Reggeization at NNLL.

We should note that the results we have presented do not imply any mis-cancellation of the
singularities of cross sections in hadron-hadron collisions, because at
the longest distance scales, which contribute to the infrared poles of
the full hadron-hadron cross section, the incoming particles are the
colourless hadrons and the colour factor associated with the
factorization violation is zero. This was a crucial step in the proof of
the factorization theorem in Ref.~\cite{Collins:1988ig}: they were able
to isolate the effect of the Coulomb/Glauber region of gluon momentum into factorization-violating terms that,
when summed over all diagrams, gave zero. Furthermore, we fully expect that the factorization breaking effects we have been discussing will cancel in inclusive observables, in accord with the theorem derived by Collins, Soper and Sterman~\cite{Collins:1988ig} and more recently by Aybat and Sterman~\cite{Aybat:2008ct}. The cancellation ought to occur upon accounting for real emissions with the same cut-diagram topology as the virtual corrections we consider here. However, these cancellations will not be complete in non-inclusive observables. In the case of super-leading logarithms, the factorization breaking is accompanied by an uncancelled soft-collinear double logarithm in an otherwise single logarithmic observable. The question of at which logarithmic order factorization breaking effects enter  has been discussed by Banfi, Salam and Zanderighi~\cite{Banfi:2010xy}. In general, we expect to be able to factorize collinear logarithms only up to a scale $\mu$ below which real emissions are summed inclusively,
and that the violations of factorization discussed here \emph{will}
contribute to the values of physical observables above that scale.

Finally, we may remark that the factorization-breaking physics we have
been exploring is not unrelated to the role of the underlying event, or
of gap survival in diffractive physics. For an
observable that is inclusive below some scale $\mu$, we may factorize
the incoming partons at that scale. The evolution in scales between
$\mu$ and the high scale of a hard process will `dress' these incoming
partons with collinear accompaniments and interactions between these
systems will violate the collinear factorization theorem, as we have
discussed. These accompanying systems will interact with each other in
an exactly analogous way to the remnants in hadron-hadron
collisions. Thus, the factorization violations discussed in this paper
will have effects that may be considered as perturbatively-calculable
contributions to the underlying event.
Furthermore, diffractive
scattering is not power suppressed and can lead to important
contributions in studies of the final state. However, it is well known
that diffractive scattering in hadron-hadron collisions does not obey a
factorization theorem: as a result one is led to the phenomenological
idea of gap survival. In the high energy limit of
scattering amplitudes, Coulomb gluon exchange is responsible for building
the perturbative manifestation of the Pomeron and the eikonal exchange
triggers the breakdown of factorization and leads to the filling of a
would-be rapidity gap; in this case the factorization violating effects
we have discussed can be thought of as the perturbative tail of the
gap survival effect.

\section*{Acknowledgements}

We are grateful to Mrinal Dasgupta and the other members of the Manchester ``QCD club'' for
extensive discussions of CdFR. Thanks also to Stefano Catani for his comments. This work was funded in part by the Lancaster-Manchester-Sheffield
Consortium for Fundamental Physics under STFC grant ST/J000418/1 and in
part (MHS) by an IPPP Associateship.


\begin{thebibliography}{99}

%\cite{Collins:1988ig}
\bibitem{Collins:1988ig}
  J.~C.~Collins, D.~E.~Soper and G.~F.~Sterman,
  ``Soft Gluons and Factorization'',
  Nucl.\ Phys.\ B {\bf 308} (1988) 833.
  %%CITATION = NUPHA,B308,833;%%

%\cite{Catani:2011st}
\bibitem{Catani:2011st}
  S.~Catani, D.~de Florian and G.~Rodrigo,
  ``Space-like (versus time-like) collinear limits in QCD: Is
  factorization violated?'',
  JHEP {\bf 1207} (2012) 026
  [arXiv:1112.4405v1 [hep-ph]].
  %%CITATION = ARXIV:1112.4405;%%

%\cite{Forshaw:2006fk}
\bibitem{Forshaw:2006fk}
  J.~R.~Forshaw, A.~Kyrieleis and M.~H.~Seymour,
  ``Super-leading logarithms in non-global observables in QCD'',
  JHEP {\bf 0608} (2006) 059
  [hep-ph/0604094].
  %%CITATION = HEP-PH/0604094;%%

%\cite{Forshaw:2008cq}
\bibitem{Forshaw:2008cq}
  J.~R.~Forshaw, A.~Kyrieleis and M.~H.~Seymour,
  ``Super-leading logarithms in non-global observables in QCD: Colour basis independent calculation'',
  JHEP {\bf 0809} (2008) 128
  [arXiv:0808.1269 [hep-ph]].
  %%CITATION = ARXIV:0808.1269;%%

%\cite{Kidonakis:1998nf}
\bibitem{Kidonakis:1998nf}
  N.~Kidonakis, G.~Oderda and G.~F.~Sterman,
  ``Evolution of color exchange in QCD hard scattering'',
  Nucl.\ Phys.\ B {\bf 531} (1998) 365
  [hep-ph/9803241].
  %%CITATION = HEP-PH/9803241;%%

%\cite{Oderda:1998en}
\bibitem{Oderda:1998en}
  G.~Oderda and G.~F.~Sterman,
  ``Energy and color flow in dijet rapidity gaps'',
  Phys.\ Rev.\ Lett.\  {\bf 81} (1998) 3591
  [hep-ph/9806530].
  %%CITATION = HEP-PH/9806530;%%

%\cite{Berger:2001ns}
\bibitem{Berger:2001ns}
  C.~F.~Berger, T.~Kucs and G.~F.~Sterman,
  ``Energy flow in interjet radiation'',
  Phys.\ Rev.\ D {\bf 65} (2002) 094031
  [hep-ph/0110004].
  %%CITATION = HEP-PH/0110004;%%

%\cite{Dasgupta:2002bw}
\bibitem{Dasgupta:2002bw}
  M.~Dasgupta and G.~P.~Salam,
  ``Accounting for coherence in interjet E(t) flow: A Case study'',
  JHEP {\bf 0203} (2002) 017
  [hep-ph/0203009].
  %%CITATION = HEP-PH/0203009;%%

%\cite{Kyrieleis:2006he}
\bibitem{conferences}
  A.~Kyrieleis,
  ``Super-leading logarithms in gaps-between-jets'',
  hep-ph/0606274,
  41st Rencontres de Moriond: QCD and Hadronic Interactions,    
  18--25 Mar.\ 2006.\\
  %%CITATION = HEP-PH/0606274;%%
%\cite{Kyrieleis:2006fh}
  A.~Kyrieleis, J.~R.~Forshaw and M.~H.~Seymour,
  ``Breakdown of QCD coherence?'',
  PoS DIFF {\bf 2006} (2006) 031
  [hep-ph/0612202],
  Diffraction 2006: 4th International Workshop on Diffraction in
  High-Energy Physics,
  5--10 Sept.\ 2006.\\
  %%CITATION = HEP-PH/0612202;%%
%\cite{Seymour:2007vw}
  M.~H.~Seymour,
  ``Breakdown of Coherence?'',
  arXiv:0710.2733 [hep-ph],
  12th International Conference on Elastic and Diffractive Scattering:
  Forward Physics and QCD,
  21--25 May 2007.\\
  %%CITATION = ARXIV:0710.2733;%%
%\cite{Forshaw:2009sf}
  J.~R.~Forshaw and M.~H.~Seymour,
  ``Soft gluons and superleading logarithms in QCD'',
  Nucl.\ Phys.\ Proc.\ Suppl.\  {\bf 191} (2009) 257
  [arXiv:0901.3037 [hep-ph]],
  Ringberg Workshop on New Trends in HERA Physics 2008,
  5--10 Oct.\ 2008.
  %%CITATION = ARXIV:0901.3037;%%

%\cite{Keates:2009dn}
\bibitem{Keates:2009dn}
  J.~Keates and M.~H.~Seymour,
  ``Super-leading logarithms in non-global observables in QCD: Fixed order calculation'',
  JHEP {\bf 0904} (2009) 040
  [arXiv:0902.0477 [hep-ph]].
  %%CITATION = ARXIV:0902.0477;%%

%\cite{Catani:1996vz}
\bibitem{Catani:1996vz}
  S.~Catani and M.~H.~Seymour,
  ``A General algorithm for calculating jet cross-sections in NLO QCD'',
  Nucl.\ Phys.\ B {\bf 485} (1997) 291
   [Erratum-ibid.\ B {\bf 510} (1998) 503]
  [hep-ph/9605323].
  %%CITATION = HEP-PH/9605323;%%

%\cite{Banfi:2010xy}
\bibitem{Banfi:2010xy}
  A.~Banfi, G.~P.~Salam and G.~Zanderighi,
  ``Phenomenology of event shapes at hadron colliders'',
  JHEP {\bf 1006} (2010) 038
  [arXiv:1001.4082 [hep-ph]].
  %%CITATION = ARXIV:1001.4082;%%

%\cite{Catani:1998bh}
\bibitem{Catani:1998bh}
  S.~Catani,
  ``The Singular behaviour of QCD amplitudes at two loop order'',
  Phys.\ Lett.\ B {\bf 427} (1998) 161
  [hep-ph/9802439].
  %%CITATION = HEP-PH/9802439;%%

%\cite{Dixon:2008gr}
\bibitem{Dixon:2008gr}
  L.~J.~Dixon, L.~Magnea and G.~F.~Sterman,
  ``Universal structure of subleading infrared poles in gauge theory amplitudes'',
  JHEP {\bf 0808} (2008) 022
  [arXiv:0805.3515 [hep-ph]].
  %%CITATION = ARXIV:0805.3515;%%

%\cite{Becher:2009cu}
\bibitem{Becher:2009cu}
  T.~Becher and M.~Neubert,
  ``Infrared singularities of scattering amplitudes in perturbative QCD'',
  Phys.\ Rev.\ Lett.\  {\bf 102} (2009) 162001
  [arXiv:0901.0722 [hep-ph]].
  %%CITATION = ARXIV:0901.0722;%%

%\cite{Gardi:2009qi}
\bibitem{Gardi:2009qi}
  E.~Gardi and L.~Magnea,
  ``Factorization constraints for soft anomalous dimensions in QCD scattering amplitudes'',
  JHEP {\bf 0903} (2009) 079
  [arXiv:0901.1091 [hep-ph]].
  %%CITATION = ARXIV:0901.1091;%%

%\cite{Becher:2009qa}
\bibitem{Becher:2009qa}
  T.~Becher and M.~Neubert,
  ``On the Structure of Infrared Singularities of Gauge-Theory Amplitudes'',
  JHEP {\bf 0906} (2009) 081
  [arXiv:0903.1126 [hep-ph]].
  %%CITATION = ARXIV:0903.1126;%%

%\cite{Dixon:2009ur}
\bibitem{Dixon:2009ur}
  L.~J.~Dixon, E.~Gardi and L.~Magnea,
  ``On soft singularities at three loops and beyond'',
  JHEP {\bf 1002} (2010) 081
  [arXiv:0910.3653 [hep-ph]].
  %%CITATION = ARXIV:0910.3653;%%

%\cite{Seymour:2005ze}
\bibitem{Seymour:2005ze}
  M.~H.~Seymour,
  ``Symmetry of anomalous dimension matrices for colour evolution of hard scattering processes'',
  JHEP {\bf 0510} (2005) 029
  [hep-ph/0508305].
  %%CITATION = HEP-PH/0508305;%%

%\cite{Seymour:2008xr}
\bibitem{Seymour:2008xr}
  M.~H.~Seymour and M.~Sj\"odahl,
  ``Symmetry of anomalous dimension matrices explained'',
  JHEP {\bf 0812} (2008) 066
  [arXiv:0810.5756 [hep-ph]].
  %%CITATION = ARXIV:0810.5756;%%

%\cite{Aybat:2008ct}
\bibitem{Aybat:2008ct}
  S.~M.~Aybat and G.~F.~Sterman,
  ``Soft-Gluon Cancellation, Phases and Factorization with Initial-State Partons'',
  Phys.\ Lett.\ B {\bf 671} (2009) 46
  [arXiv:0811.0246 [hep-ph]].
  %%CITATION = ARXIV:0811.0246;%%

%\cite{DelDuca:2011ae}
\bibitem{DelDuca:2011ae}
  V.~Del Duca, C.~Duhr, E.~Gardi, L.~Magnea and C.~D.~White,
  ``The Infrared structure of gauge theory amplitudes in the high-energy limit'',
  JHEP {\bf 1112} (2011) 021
  [arXiv:1109.3581 [hep-ph]].
  %%CITATION = ARXIV:1109.3581;%%

\end{thebibliography}
\end{document}